# Towards an Intelligent Edge: Wireless Communication Meets Machine Learning

Guangxu Zhu[1], Dongzhu Liu[1], Yuqing Du[1], Changsheng You[1], Jun Zhang[2] and Kaibin Huang[1]

## Abstract

The recent revival of artificial intelligence (AI) is revolutionizing almost every branch of science and technology. Given the ubiquitous smart mobile gadgets and *Internet of Things* (IoT) devices, it is expected that a majority of intelligent applications will be deployed at the edge of wireless networks. This trend has generated strong interests in realizing an "intelligent edge" to support AI-enabled applications at various edge devices. Accordingly, a new research area, called *edge learning,* emerges, which crosses and revolutionizes two disciplines: wireless communication and machine learning. A major theme in edge learning is to overcome the limited computing power, as well as limited data, at each edge device. This is accomplished by leveraging the *mobile edge computing* (MEC) platform and exploiting the massive data distributed over a large number of edge devices. In such systems, learning from distributed data and communicating between the edge server and devices are two critical and coupled aspects, and their fusion poses many new research challenges. This article advocates a new set of design principles for wireless communication in edge learning, collectively called *learning-driven communication.* Illustrative examples are provided to demonstrate the effectiveness of these design principles, and unique research opportunities are identified.

## I. Introduction

We are witnessing a phenomenal growth in global data traffic, accelerated by the increasing popularity of mobile devices, e.g., smartphones, tablets and sensors. According to the *intersectional data corporation* (IDC), there will be 80 billion devices connected to the Internet by 2025, and the global data will reach 163 zettabytes, which is ten times of the data generated in 2016 [1]. The unprecedented amount of data, together with the recent breakthroughs in *artificial intelligence* (AI), inspire people to envision ubiquitous computing and ambient intelligence, which will not only improve our life qualities but also provide a platform for scientific discoveries and engineering innovations. In particular, this vision is driving the industry and academia to vehemently invest in technologies for creating an *intelligent (network) edge,* which supports emerging application scenarios such as smart city, eHealth, eBanking, intelligent transportation, etc. This has led to the emergence of a new research area, called *edge learning,* which refers to the deployment of machine-learning algorithms at the network edge [2]. The key motivation of pushing learning towards the edge is to allow rapid access to the enormous real-time data generated by the edge devices for fast AI-model training, which in turn endows on the devices human-like intelligence to respond to real-time events.

Traditionally, training an AI model, especially a deep model, is computation-intensive and thus can only be supported at powerful cloud servers. Riding the recent trend in developing the *mobile edge computing* (MEC) platform, training an AI model is no longer exclusive for cloud servers but also affordable at edge servers. Particularly, the network virtualization architecture recently standardized by 3GPP is able to support edge learning on top of edge computing [3]. Moreover, the latest mobile devices are also armed with high-performance *central-processing units* (CPUs) or *graphics processing units* (GPUs) (e.g., A11 bionic chip in iPhone X), making them capable in training some small-scale AI models. The coexistence of cloud, edge and on-device learning paradigms has led to a layered architecture for in-network machine learning, as shown in Fig. 1. Different layers possess different data processing and storage capabilities, and cater for different types of learning applications with distinct latency and bandwidth requirements.

G. Zhu, D. Liu, Y, Du, C, You and K. Huang are with the Dept. of Electrical and Electronic Engineering at the University of Hong Kong, Hong Kong. Email: {gxzhu,dzliu,yqdu,csyou,huangkb}@eee.hku.hk. Corresponding Author: K. Huang.

J. Zhang is with the Dept. of Electronic and Computer Engineering at the Hong Kong University of Science and Technology, Hong Kong. Email: eejzhang@ust.hk

Compared with cloud and on-device learning, edge learning has its unique strengths. First, it has the most balanced resource support (see Fig. 1), which helps achieving the best tradeoff between the AI-model complexity and the model-training speed. Second, given its proximity to data sources, edge learning overcomes the drawback of cloud learning that fails to process real-time data due to excessive propagation delay and also network congestion caused by uploading data to the cloud. Furthermore, the proximity gives an additional advantage of location-and-context awareness. Last, compared with on-device learning, edge learning achieves much higher learning accuracy by supporting more complex models and more importantly aggregating distributed data from many devices. Due to the all-rounded capabilities, edge learning can support a wide spectrum of AI models to power a broad range of mission-critical applications, such as auto-driving, rescue-operation robots, disaster avoidance and fast industrial control. Nevertheless, edge learning is at its nascent stage and thus remains a largely uncharted area with many open challenges.

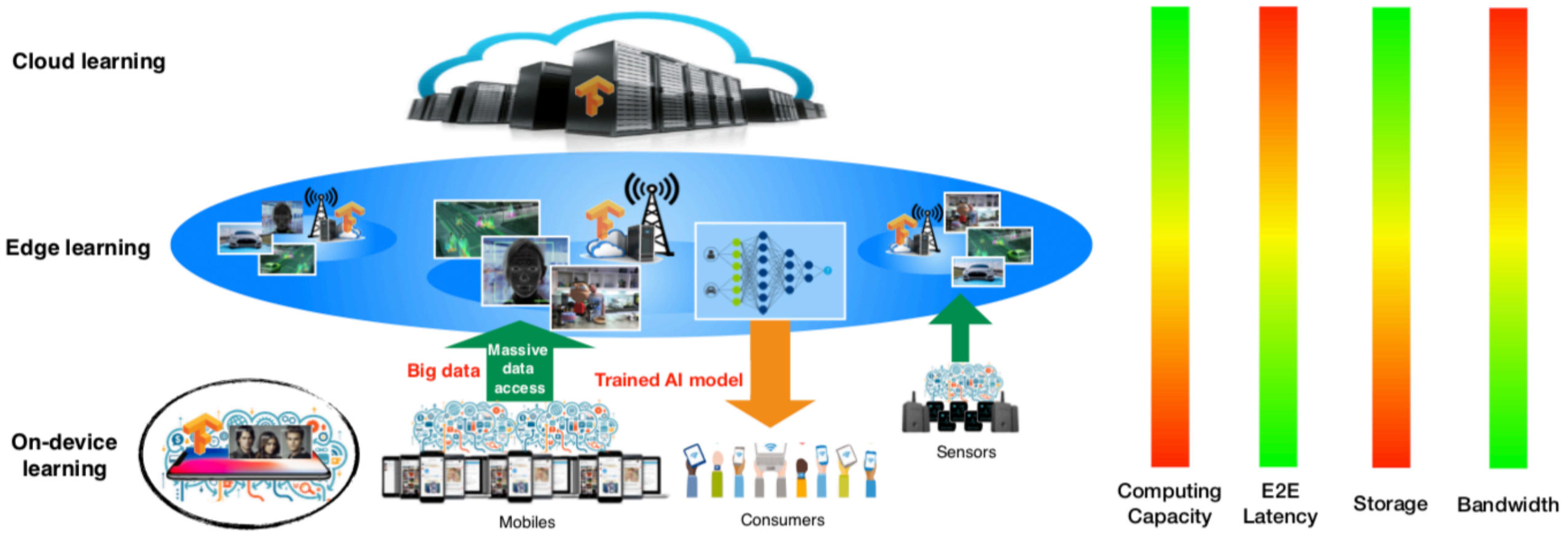

Fig. 1. Layered in-network machine learning architecture.

The main design objective in edge learning is the *fast intelligence acquisition* from the rich but highly distributed data at subscribed edge devices. This critically depends on data processing at edge servers, as well as efficient communication between edge servers and edge devices. Compared with increasingly high processing speeds at edge servers, communication suffers from hostility of wireless channels (e.g., pathloss, shadowing, and fading), and consequently forms the bottleneck for ultra-fast edge learning. In order to distill the shared intelligence from distributed data, excessive communication latency may arise from the need of uploading to an edge server a vast amount of data generated by *millions to billions* of edge devices, as illustrated in Fig. 1. As a concrete example, the Tesla's AI model for auto-driving is continuously improved using RADAR and LIDAR sensing data uploaded by millions of Tesla vehicles on the road, which can amount to about 4,000 GB for one car per day. Given the enormity in data and the scarcity of radio resources, **how to fully exploit the distributed data in AI-model training without incurring excessive communication latency** poses a grand challenge for *wireless data acquisition* in edge learning.

Unfortunately, the state-of-the-art wireless technologies are incapable of tackling the challenge. The fundamental reason is that the traditional design objectives of wireless communications, namely communication reliability and data-rate maximization, do not directly match that of edge learning. This means that we have to break away from the conventional philosophy in traditional wireless communication, which can be regarded as a "*communication-computing separation*" approach. Instead, we should exploit the coupling between communication and learning in edge learning systems. To materialise the new philosophy, we propose in this article a set of new design principles for wireless communication in edge learning, collectively called *learning-driven communication*. In the following sections, we shall discuss specific research directions and provide concrete examples to illustrate this paradigm shift, which cover key communication aspects including multiple access, resource allocation and signal encoding, as summarized in Table 1. All of these new design principles share a common principle as highlighted below.

**Principle of Learning-Driven Communication - Fast Intelligence Acquisition**

Efficiently transmit data or learning-relevant information to *speed up and improve AI-model training* at edge servers.

*Table 1. Conventional Communication versus Learning-Driven Communication*

| Commun. Tech. | Item | Conventional Commun. | Learning-Driven Commun. |
|---|---|---|---|
| **Multiple access (Section II)** | Target | Decoupling messages from users | Computing func. of distributed data |
| | Case study | OFDMA | Model-update averaging by AirComp |
| **Resource Allocation (Section III)** | Target | Maximize sum-rate or reliability | Fast intelligence acquisition |
| | Case study | Reliability-based retransmission | Importance-aware retransmission |
| **Signal Encoding (Section IV)** | Target | Optimal tradeoffs between rate and distortion/reliability | Latency minimization while preserving the learning accuracy |
| | Case study | Quantization, adaptive modulation and polar code | Grassmann analog encoding |

At the high level, learning-driven communication integrates wireless communication and machine leaning that have been rapidly advancing as two separate disciplines with few cross-paths. In this paper, we aim at providing a roadmap for this emerging and exciting area by highlighting research opportunities, shedding light on potential solutions, as well as discussing implementation issues.

## II. Learning-Driven Multiple Access

### a) Motivation and Principle

In edge learning, the involved training data (e.g., photos, social-networking records, and user-behaviour data) are often privacy sensitive and large in quantity. Thus uploading them from devices to an edge server for centralized model training may not only raise a privacy concern but also incur prohibitive cost in communication. This motivates an innovative edge-learning framework, called *federated learning*, which features *distributed learning* at edge devices and *model-update aggregation* at an edge server [4]. Federated learning can effectively address the aforementioned issues as only the locally computed model updates, instead of raw data, are uploaded to the server. A typical federated-learning algorithm alternates between two phases, as shown in Fig. 2 (a). One is to aggregate distributed model updates over a multi-access channel and apply their average to update the AI-model at the edge server. The other is to broadcast the model under training to allow edge devices to continuously refine their individual versions of the model. This learning framework is used as a particular scenario of edge learning in this section to illustrate the new design principle of learning-driven multiple access.

Model-update uploading in federated learning is bandwidth-consuming as an AI model usually comprises millions to billions of parameters. Overall the model updates by thousands of edge devices may easily congest the air-interface, making it a bottleneck for agile edge learning. The said bottleneck is arguably an artifact of the classic approach of *communication-computing separation*. Existing multiple access technologies such as *orthogonal frequency-division multiple access* (OFDMA) and *code division multiple access* (CDMA) are purely for rate-driven communication and fail to adapt to the actual learning task. The need for enabling fast edge learning from massive distributed data calls for a new design principle for multiple access. In this section, we present *learning-driven multiple access* as the solution, and showcase a particular technique under this new principle.

The key innovation underpinning the learning-driven multiple access is to exploit the insight that the learning task involves computing some aggregating function (e.g., averaging or finding the maximum) of multiple data samples, rather than decoding individual samples as in the existing scheme. For example, in federated learning, the edge server requires the average of model updates rather than their individual values. On the other hand, the multi-access wireless channel by itself is a natural data aggregator: the simultaneously transmitted analog-waves by different devices are automatically superposed at the receiver but weighed by the channel coefficients. The above insights motivate the following design principle for multiple access in edge learning. It changes the traditional philosophy of "overcoming interference" to the new one of "harnessing interference".

**Principle of Learning-Driven Multiple Access**

Unique characteristics of wireless channels, such as broadcast and superposition, should be exploited for functional computation over distributed data to accelerate edge learning.

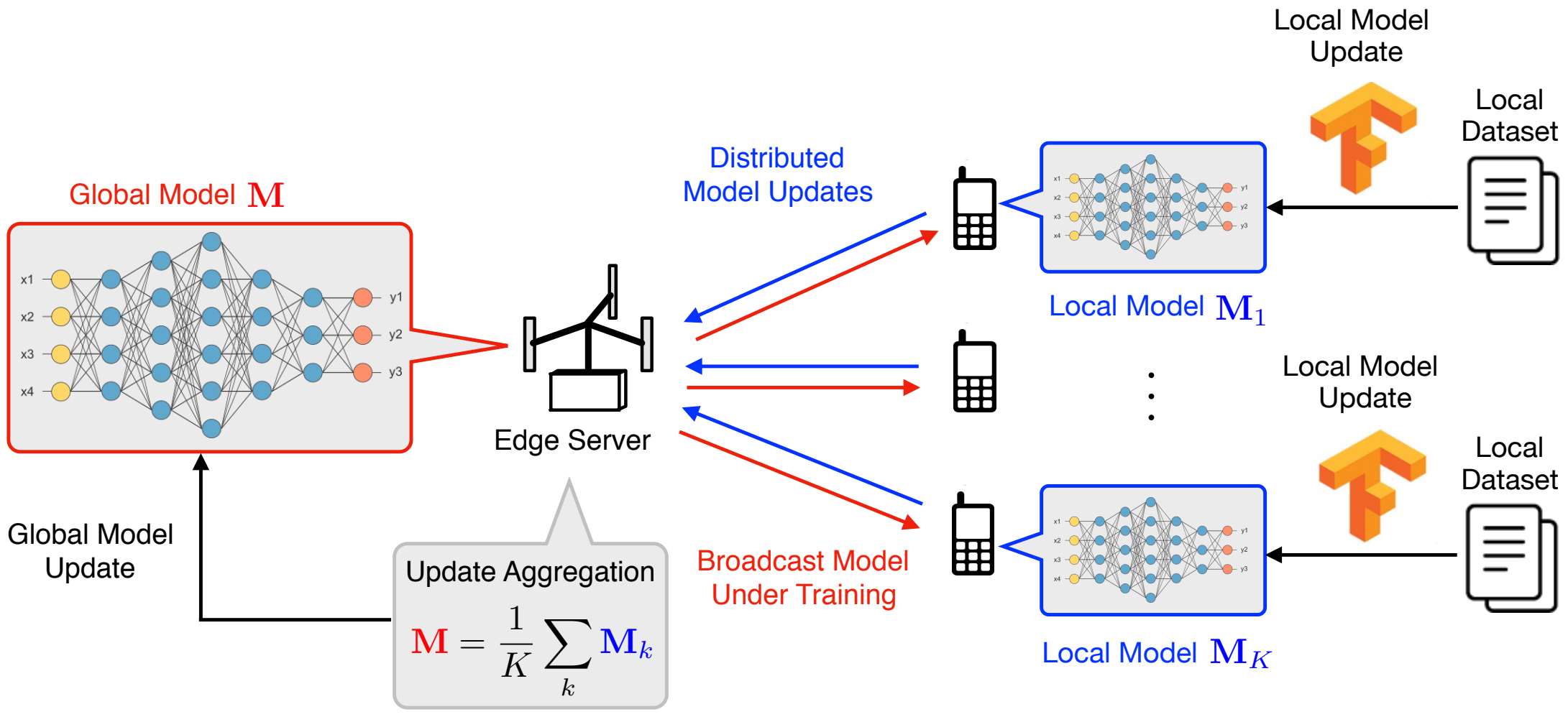

(a)

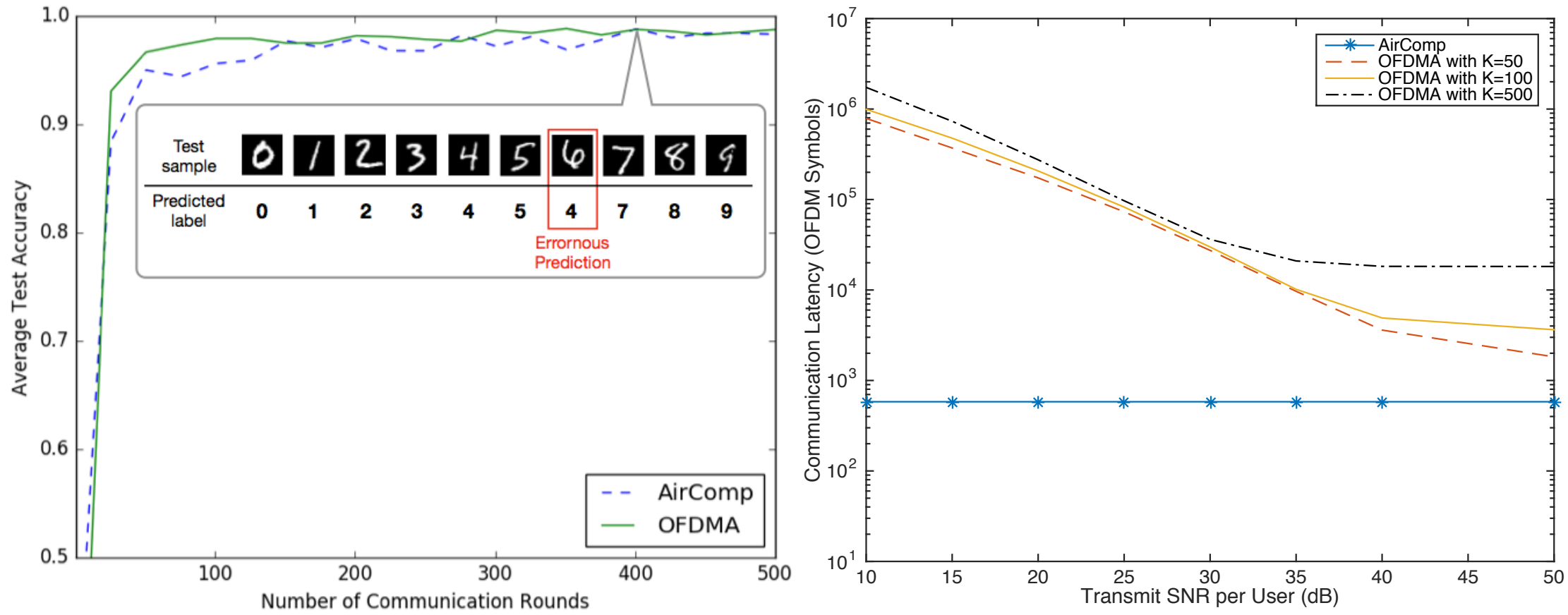

(b)

Fig. 2. (a) Federated learning using wirelessly distributed data. (b) Performance comparison between AirComp and OFDMA in test accuracy (left) and communication latency (right). The implementation details are specified as follows. For AirComp, model parameters are analog-modulated and each sub-band is dedicated for single-parameter transmission; truncated-channel inversion (power control) under the transmit-power constraint is used to tackle the channel fading. For OFDMA, model parameters are first quantized into a bit sequence (16-bit per parameter). Then adaptive MQAM modulation is adopted to adapt the data rate to the channel condition such that the spectrum efficiency is maximized while the target bit-error-rate of $10^{-3}$ is maintained.

Following the new principle, the superposition nature of the multi-access channel suggests that by using linear-analog modulation and pre-channel-compensation at the transmitter, the "interference" caused by concurrent data transmission can be exploited for fast data aggregation. This intuition has been captured by a recently proposed technique called *over-the-air computation* (AirComp) [5], [6]. By allowing simultaneous transmission, AirComp can dramatically reduce the multiple access latency by a factor equal to the number of users (i.e., 100 times for 100 users). It provides a promising solution for overcoming the communication latency bottleneck in edge learning.

### b) Case Study: Over-the-Air Computation for Federated Learning

**Experiment settings:** Consider a federated learning system with one edge server and $K$=100 edge devices. For exposition, we consider the learning task of handwritten-digit recognition using the well-known MNIST dataset that consists of 10 categories ranging from digit "0" to "9" and a total of 60000 labeled training data samples. To simulate the distributed mobile data, we randomly partition the training samples into 100 equal shares, each of which is assigned to one particular device. The classifier model is implemented using a 4-layer *convolutional neural network* (CNN) with two 5x5 convolution layers, a fully connected layer with 512 units and ReLu activation, and a final softmax output layer (582,026 parameters in total).

**AirComp versus OFDMA**: During the federated model training, in each *communication round*, local models trained at edge devices (using e.g., stochastic gradient descend) are transmitted and aggregated at the edge server over a shared broadband channel that consists of $N_s$=1000 orthogonal sub-channels. Two multiple access schemes, namely the conventional OFDMA and the proposed AirComp, are compared. They mainly differ in how the available sub-channels are shared. For OFDMA, the 1000 sub-channels are evenly allocated to the $K$ edge devices, so each device uploads its local model using only fractional bandwidth that reduces as $K$ grows. Model averaging is performed by the edge server after all local models are reliably received, and thus the communication latency is determined by the slowest device. In contrast, the AirComp scheme allows every device to use the full bandwidth so as to exploit the "interference" for direct model averaging over the air. The latency of AirComp is thus independent of the number of accessing devices.

**Performance**: The learning accuracy and communication latency of the two schemes are compared in Fig. 2 (b) under the same transmit *signal-to-noise ratio* (SNR) per user. As shown at the left-hand side of Fig. 2 (b), although AirComp is expected to be more vulnerable to channel noise, it is interesting to see that the two schemes are comparable in learning accuracy. Such accurate learning of AirComp is partly due to the high expressiveness of the deep neural network which makes the learnt model robust against perturbation by channel noise. The result has a profound and refreshing implication that *reliable communication may not be the primary concern in edge learning*. Essentially, AirComp exploits this relaxation on communication reliability to trade for a low communication latency as shown at the right-hand side of Fig. 2 (b). The latency gap between the two schemes is remarkable. Without compromising the learning accuracy, AirComp can achieve a significant latency reduction ranging from 10x to 1000x. In general, the superiority in latency of AirComp over OFDMA is more pronounced in the low SNR regime and dense-network scenarios.

### c) Research Opportunities

The new design principle of learning-driven multiple access points to numerous research opportunities, some of which are described as follows.

- **Robust learning with imperfect AirComp**. Wireless data aggregation via AirComp requires channel pre-equalization at the transmitting devices. Inaccurate channel estimation and non-ideal hardware at the low-cost edge devices may cause imperfect equalization and thus distort the aggregated data. For practical implementation, it is important to characterize the effects of the imperfect AirComp on the performance of edge learning, based on which new techniques can be designed to improve the learning robustness.
- **Asynchronous AirComp**. Successful implementation of AirComp requires strict synchronization between all the participating edge devices. This may be hard to achieve when the devices exhibit high mobility or the learning system is highly dynamic with the participating devices changing frequently over time. To enable ultra-fast data aggregation in these scenarios, new schemes operated in an asynchronous manner or with a relaxed requirement on synchronization are desirable.

- **Generalization to other edge-learning architectures**. The proposed AirComp solution targets federated-learning architecture. It may not be applicable for other architectures where the edge server needs to perform more sophisticated computation over the received data than simple averaging. How to exploit the superposition property of a multi-access channel to compute more complex functions is the main challenge in generalizing the current learning-driven multiple access design to other architectures.

## III. Learning-Driven Radio Resource Management

### a) Motivation and Principle

Based on the traditional approach of communication-computing separation, existing methods of *radio-resource management* (RRM) are designed to maximize the efficiency of spectrum utilization by carefully allocating the scarce radio resources such as power, frequency band and access time. However, such an approach is no longer effective in edge learning, as it fails to exploit the subsequent learning process for further performance improvement. This motivates us to propose the following design principle for RRM in edge learning.

<u>**Principle of Learning-Driven RRM**</u>

Radio resources should be allocated based on the value of transmitted data so as to optimize the edge-learning performance.

Conventional RRM assumes that the transmitted messages have the same value for the receiver. The assumption makes sum-rate maximization a key design criterion. When it comes to edge learning, the rate-driven approach is no longer efficient as some messages tend to be more valuable than others for training an AI model.

In this part, we introduce a representative technique following the above learning-driven design principle, called *importance-aware resource allocation,* which takes the data importance into account in resource allocation. The basic idea of this new technique shares some similarity with a key area in machine learning called *active learning*. Principally, active learning is to select important samples from a large unlabelled dataset for labelling (by querying an oracle) so as to accelerate model training with a labelling budget [7]. A widely adopted measure of (data) importance is uncertainty. To be specific, a data sample is more uncertain if it is less confidently predicted by the current model. For example, a cat photo that is classified as "cat" with a correctness probability of 0.6 is more uncertain than that of a probability of 0.8. A commonly used uncertainty measure is entropy, a notion from information theory. As its evaluation is complex, a heuristic but simple alternative is the distance of a data sample from the decision boundaries of the current model. Taking *support vector machine* (SVM) as an example, a training data sample near to the decision boundary is likely to become a *support vector*, thereby contributing to defining the classifier. In contrast, a sample away from boundaries makes no such contribution.

Compared with active learning, learning-driven RRM has its additional challenges given the volatile wireless channels. In particular, besides data importance, it needs to consider radio-resource allocation to ensure a certain level of reliability in transmitting a data sample. A basic diagram of learning-driven RRM is illustrated in Fig.3 (a). In the following, we will provide a concrete case-study for illustration.

### b) Case Study: Importance-Aware Retransmission for Wireless Data Acquisition

<u>**Experiment settings:**</u> Consider an edge learning system where a classifier is trained at the edge server based on SVM, with data collected from distributed edge devices. The acquisition of high-dimensional training-data samples is bandwidth consuming and relies on a noisy data channel. On the other hand, a low-rate reliable channel is allocated for accurately transmitting small-size labels. The mismatch between the labels and noisy data samples at the edge server may lead to an incorrectly learnt model. To tackle the issue, importance-aware retransmission with coherent combining is used to enhance the data quality. The radio resource is specified by the limited transmission budgets with $N$=4000 samples (new/retransmitted). To train the classifier, we use the MNIST dataset described in Section II-b) and choose the relatively less differentiable class pair of '3' and '5' to focus on a binary classification case.

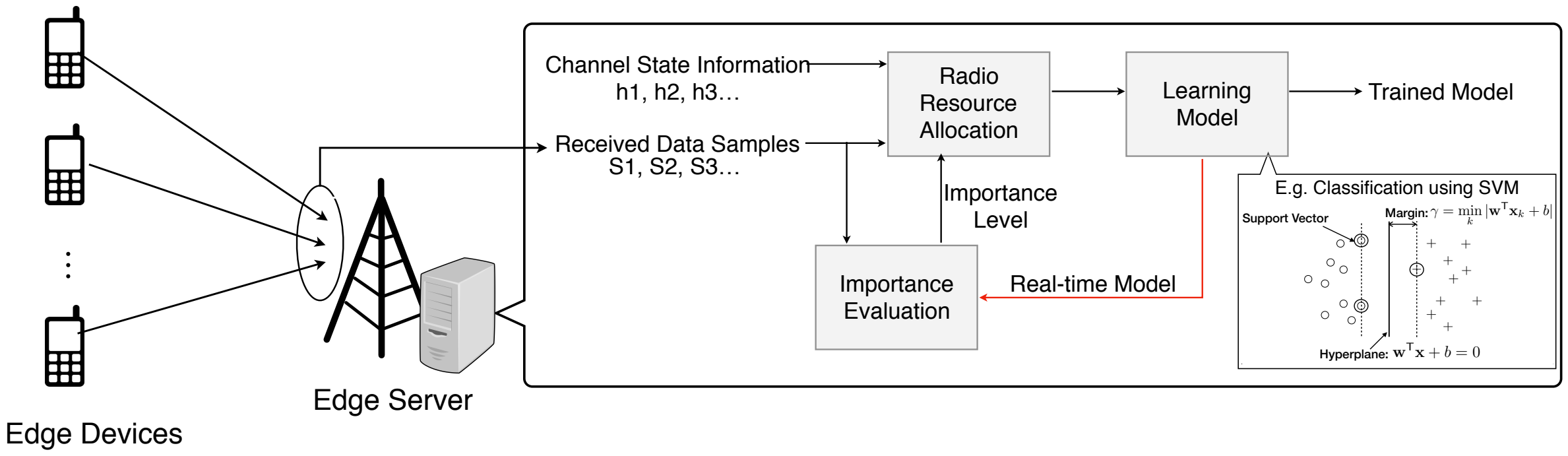

(a)

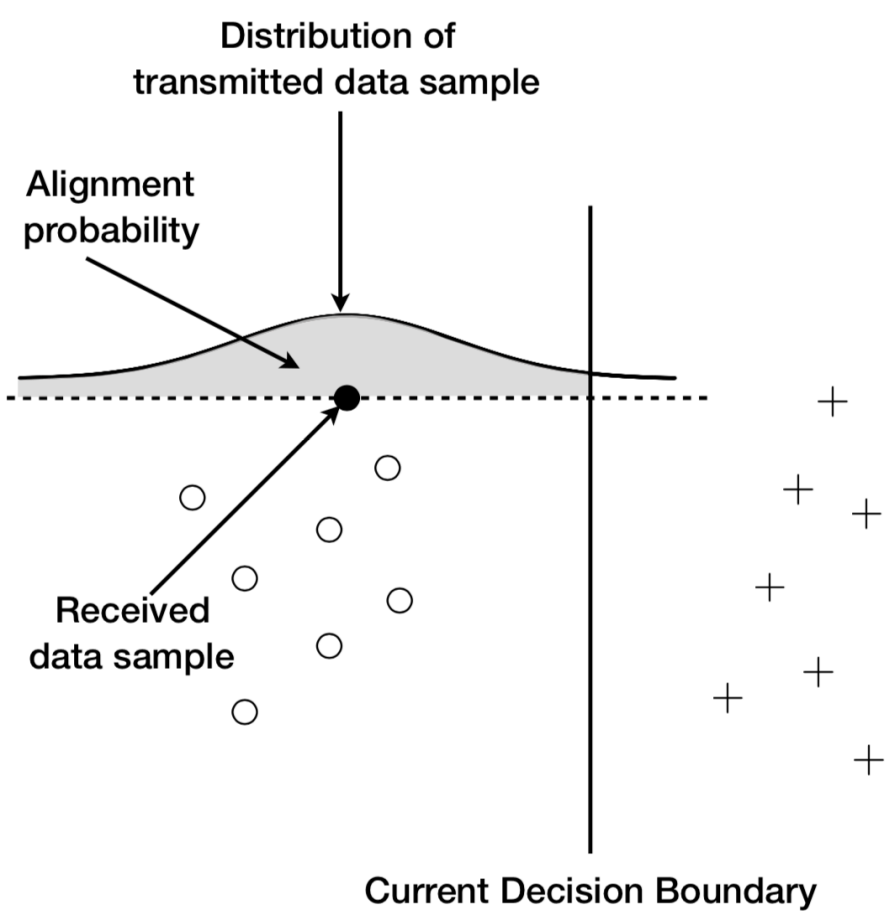

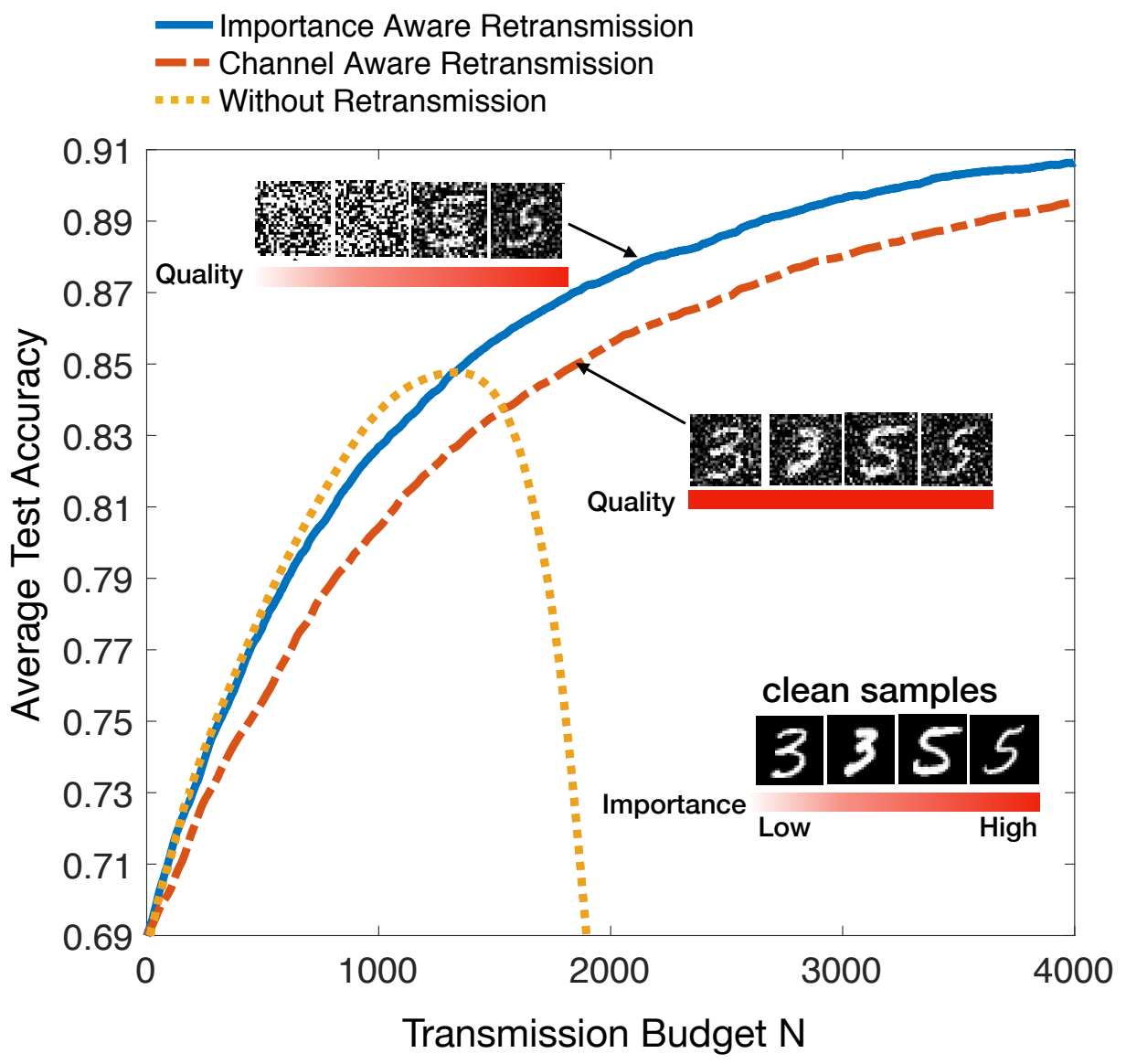

(b) (c)

Fig. 3 (a) A communication system with learning-driven RRM. (b) Illustration of the issue of data-label mismatch which is equivalent to that transmitted and received samples lie at different sides of the decision boundary (misalignment). (c) Classification performance for importance-aware retransmission and two baselines. The MRC combining technique is applied to coherently combine all the retransmission observations for maximizing the receive SNR. The retransmission stops when the receive SNR meets a predefined threshold. The average receive SNR is 10 dB.

**Importance-aware retransmission:** Under a transmission budget constraint, the RRM problem can be specified as: How many retransmission instances should be allocated for a given data sample? Concretely, in each communication round, the edge server should make a binary decision on either selecting a device for acquiring a new sample or requesting the previously scheduled device for retransmission to improve sample quality. Given a finite transmission budget, the decision making needs to address the *tradeoff between the quality and quantity* of received samples. As shown in Fig. 3 (b), data located near to the decision boundary are more critical to the model training but also easier to commit the data-label mismatch issue. Therefore , they require more retransmission budget to ensure a pre-specified alignment probability (defined as the possibility that the transmitted and received data lie at the same side of the decision boundary). This motivates the said importance-aware retransmission scheme where the retransmission decision making is controlled by applying an adaptive SNR threshold. The adaptation is realized by weighting the threshold with a coefficient that is equal to the distance between the sample to the decision boundary. This enables an intelligent allocation of the transmission budgets according to the data importance such that the optimal quality-quantity tradeoff can be achieved.

**Performance:** Fig. 3 (c) presents the learning performance of the importance-aware retransmission along with two benchmark schemes, namely, conventional channel-aware retransmission with a fixed SNR threshold and the scheme without retransmission. It is observed that if there is no retransmission, the learning performance dramatically degrades after acquiring a sufficiently large number of noisy samples. This is because the strong noise effect accumulates to cause the divergence of the model which justifies the need for retransmission. Next, one can observe that importance-aware retransmission outperforms the conventional channel-aware retransmission throughout the entire training duration. This confirms the performance gain from intelligent utilization of the radio source for data acquisition. The effect of importance-aware resource allocation can be further visualized by the selected four training samples as shown in the figure: the quality varies with data importance in the proposed scheme while the conventional channel-aware scheme strives to keep high quality for each data sample. This illustrates the proposed design principle and shows its effectiveness in adapting retransmission to data importance.

### c) Research Opportunities

Effective RRM plays an important role in edge learning, and the learning-driven design principle presented above brings many interesting research opportunities. A few are described below.

- **Cache-assisted importance-aware RRM**. The performance of the developed importance-aware RRM can be further enhanced by exploiting the storage of edge devices. With sufficient storage space, edge devices may pre-select important data from the locally cached data before uploading, which can result in faster convergence of the AI-model training. However, the conventional importance evaluation based on data uncertainty may lead to undesired selection of outliners. How to incorporate the data representativeness into the data importance evaluation by intelligently exploiting the distribution of local dataset is the key issue to be addressed.
- **Multi-user RRM for faster intelligence acquisition**. Multiple access technologies such as OFDMA allow simultaneous data uploading from multiple users. The resultant batch data acquisition has an obvious advantage in enhancing overall efficiency. This mainly results from the fact that the batch-data processing reduces the frequency of updating an AI-model under training. However, due to the correlation of data across different users, accelerating model training may be at a cost of unnecessarily processing redundant information that has little contribution to improving the learning performance. Therefore, how to efficiently exploit data diversity in the presence of inter-user correlation is an interesting topic to be investigated on multi-user RRM design.
- **Learning-driven RRM in diversified scenarios**. In the case-study presented above, importance-aware RRM assumes the need of uploading raw data. However, in a more general edge-learning system, what is uploaded from edge devices to the edge server is not necessarily the original data samples but can be other learning-related contents (e.g., model updates in the federated learning presented in Section II). This makes the presented data-importance-aware RRM design not directly applicable to these scenarios. As a result, a set of learning-driven RRM designs should be proposed targeting different edge-learning systems.

## IV. Learning-Driven Signal Encoding

### a) Motivation and Principle

In machine learning, feature-extraction techniques are widely applied in pre-processing raw data so as to reduce its dimensions as well as improving the learning performance. There exist numerous feature-extraction techniques. For the regression task, *principal component analysis* (PCA) is a popular technique for identifying a latent feature space and using it to reduce data samples to their low-dimensional features essential for training a model. Thereby, model overfitting is avoided. On the other hand, *linear discriminant analysis* (LDA) finds the most discriminant feature space to facilitate data classification. Moreover, *independent component analysis* (ICA) identifies the independent features of a multivariate signal which finds applications such as blind source separation. A common theme shared by feature-extraction techniques is to reduce a training dataset into low-dimensional features that simplify learning and improve its performance. In the feature-extraction process, too aggressive and too conservative dimensionality-reduction can both degrade the learning performance. Furthermore, the choice of a feature space directly affects the performance of a targeted learning task. These make designing feature-extraction techniques a challenging but important topic in machine learning.

In wireless communication, techniques of source-and-channel encoding are developed to also "preprocess" transmitted data but for a different purpose, namely efficient-and-reliable delivery. Source coding samples, quantizes, and compresses the source signal such that it can be represented by a minimum number of bits under a constraint on signal distortion. This gives rise to a *rate-distortion tradeoff.* On the other hand, for reliable transmission, channel coding introduces redundancy into a transmitted signal so as to protect it against noise and hostility of wireless channels. This results in the *rate-reliability tradeoff.* Designing joint source-and-channel coding essentially involves the joint optimization of the two mentioned tradeoffs.

Since both are data-preprocessing operations, it is natural to integrate feature extraction and source-and-channel encoding so as to enable efficient communication and learning in edge-learning systems. This gives rise to the area of *learning-driven signal encoding* with the following design principle.

**Principle of Learning-Driven Signal Encoding**

Signal encoding at an edge device should be designed by jointly optimizing feature extraction, source coding, and channel encoding so as to accelerate edge learning.

In this section, an example technique following the above principle, called *Grassmann analog encoding* (GAE), is discussed. GAE represents a raw data sample in the Euclidean space by a subspace, which can be interpreted as a feature, via projecting the sample onto a Grassmann manifold (a space of subspaces). An example is illustrated in Fig. 4 (a), where data samples in the 3-dimensional Euclidean space are projected on the Grassmann manifold. The operation reduces the data dimensionality but as a result distorts the data sample by causing *degree-of-freedom* (DoF) loss. In return, the direct transmission of GAE encoded data samples (subspaces) using linear-analog modulation not only supports blind *multiple-input-multiple-output* (MIMO) transmission without *channel-state information* (CSI) but also provides robustness against fast fading. The feasibility of blind transmission is due to the same principle as the classic non-coherent MIMO transmission [8]. On the other hand, the GAE encoded dataset retains its original cluster structure and thus its usefulness for training a classifier at the edge server. The GAE design represents an initial step towards learning-driving signal encoding for fast edge learning.

### b) Case Study: Fast Analog Transmission and Grassmann Learning

**Experiment settings:** Consider an edge-learning system, where an edge server trains a classifier using a training dataset transmitted by multiple edge devices with high mobility. The transmissions by devices are based on time sharing and independent of channels given no CSI. All nodes are equipped with antenna arrays, resulting in a set of narrow-band MIMO channels. In this case study, we focus on transmission of data samples that dominates the data acquisition process. Similar as Section III.b), labels are transmitted over a low-rate noiseless *label channel*. The data samples at different edge devices are assumed to be *independent and identically distributed* (i.i.d.) based on the classic *mixture of Gaussian* (MoG) model. The number of classes is C = 2, and each data sample is a 1×48 vector. The temporal correction of each 4×2 MIMO channel follows the classic Clark's model based on the assumption of rich scattering, where the channel-variation speed is specified by the normalized Doppler shift $f_D T_s = 0.01$, with $f_D$ and $T_s$ denoting the Doppler shift and the baseband sampling interval (or time slot), respectively. The training and test datasets are generated based on the discussed MoG model, which comprise 200 and 2000 samples, respectively. After detecting the GAE encoded data at the edge side, the Bayesian classifier [8] on the Grassmann manifold is trained for data labelling.

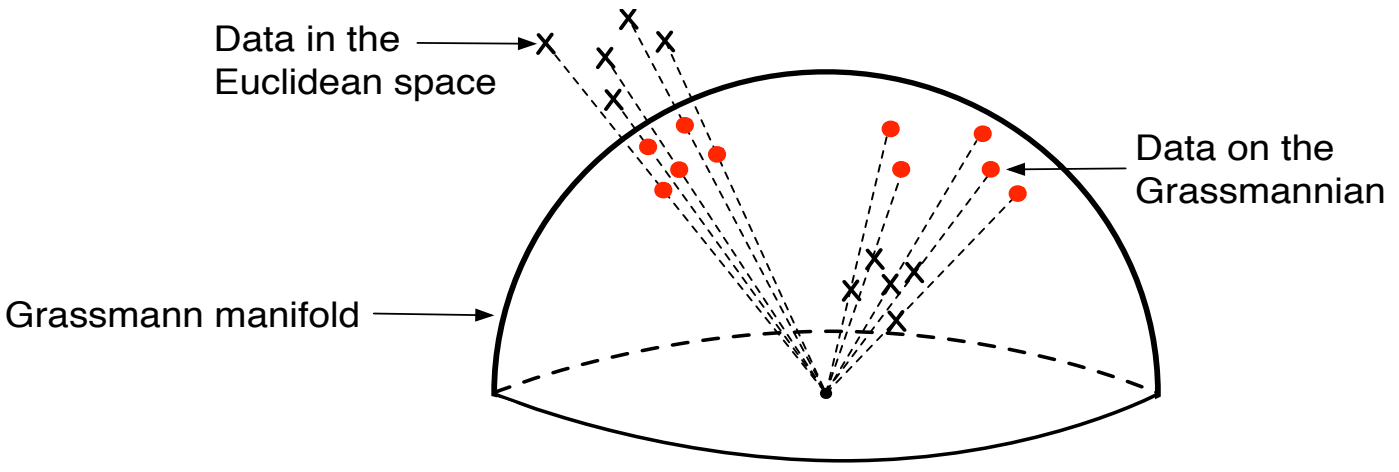

(a)

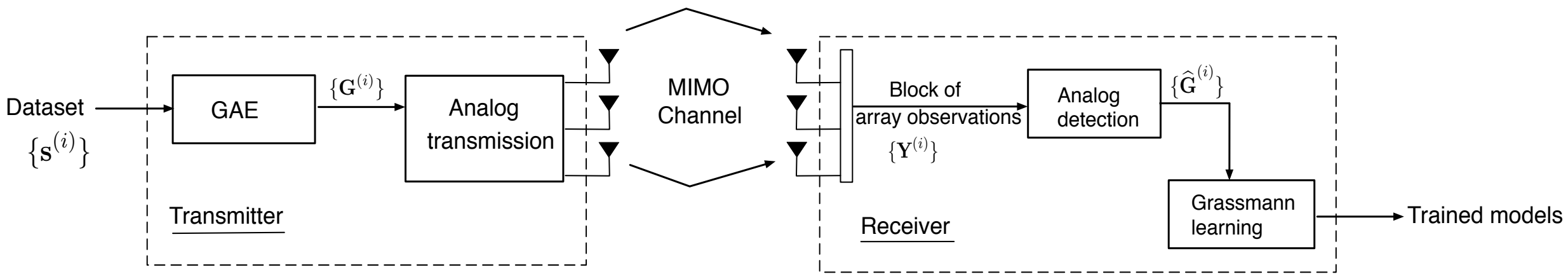

(b)

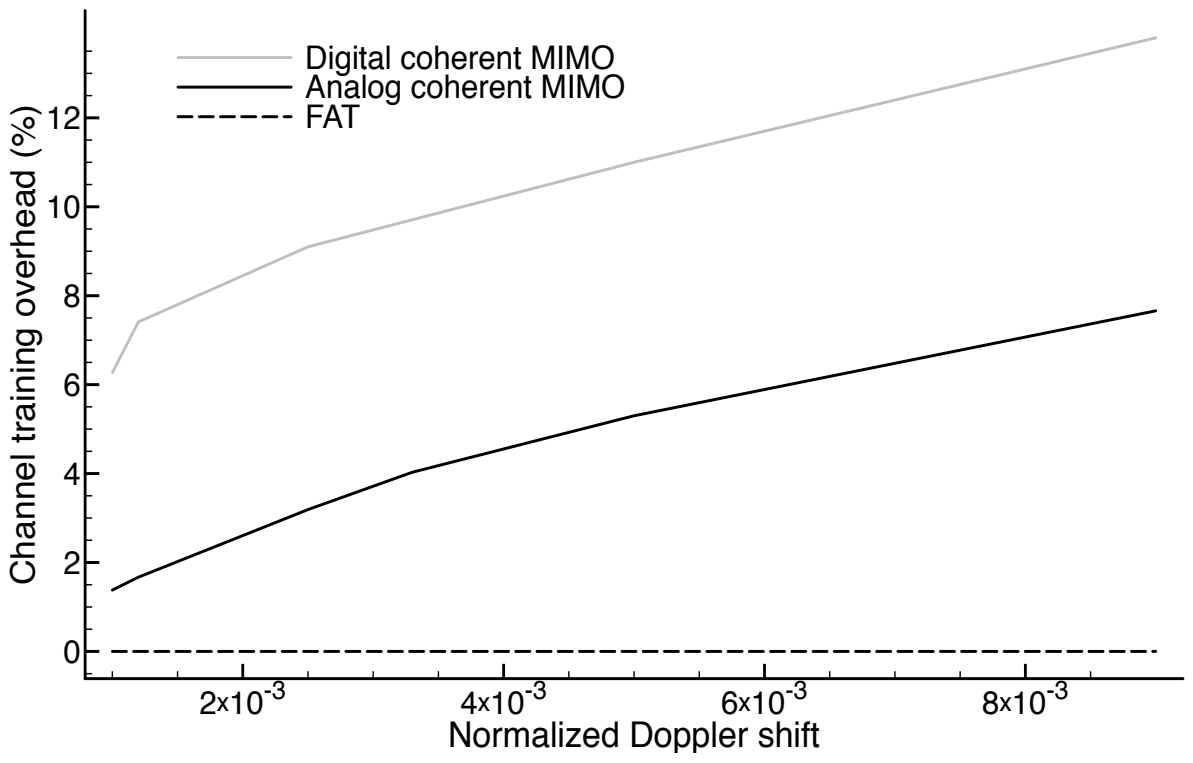

(c)

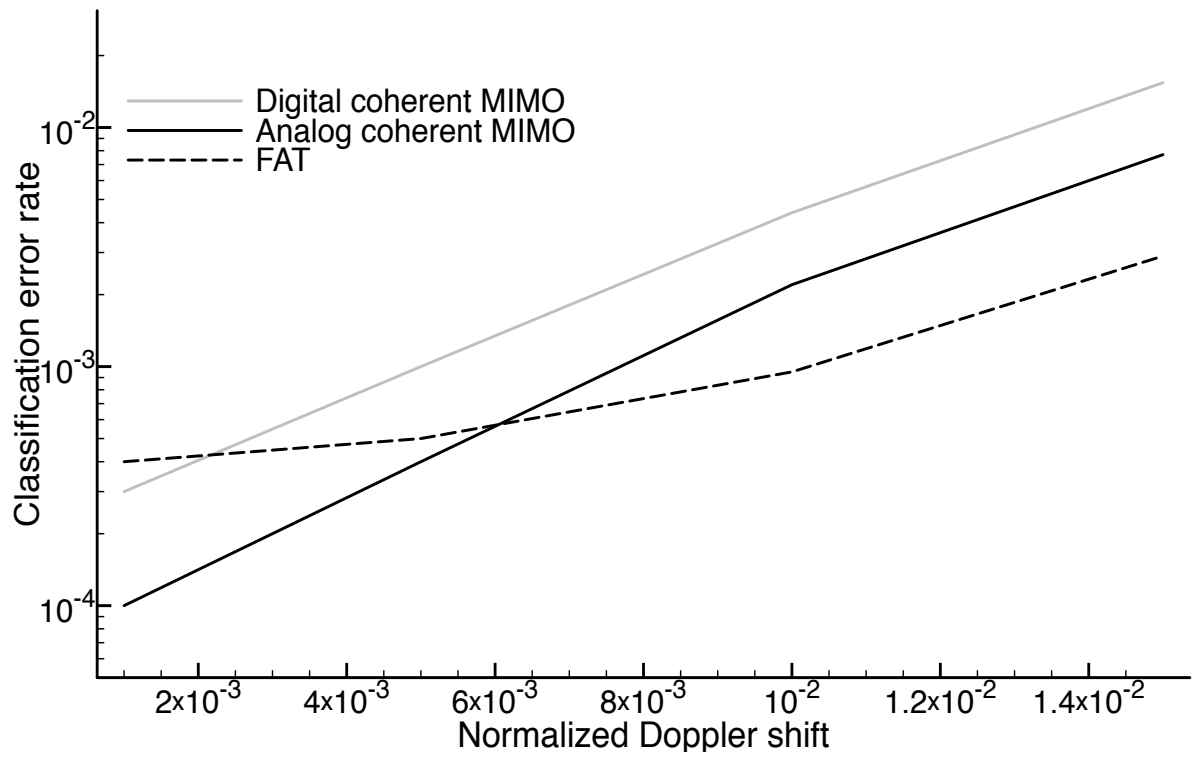

(d)

Fig. 4 (a) Principle of Grassmann analog encoding. (b) An edge-learning system based on FAT. (c) The channel-training overhead versus normalized Doppler shift for the target classification error rate of $10^{-3}$. (d) Effect of Doppler shift. The implementation details are specified as follows. Like FAT, analog MIMO transmits data samples directly by linear-analog modulation but without GAE, thus requiring channel training. On the other hand, digital MIMO quantizes data samples into 8-bit per coefficient and modulates each symbol using QPSK before MIMO transmission. All considered schemes have no error control coding.

**FAT versus coherent schemes**: A novel design, called *fast analog transmission* (FAT), having the mentioned GAE as its key component, is proposed in [8] for fast edge classification, as illustrated in Fig. 4 (b). The performance of FAT is benchmarked against two high-rate coherent schemes: digital and analog MIMO transmission, both of which assume an MMSE linear receiver and thus require channel training to acquire the needed CSI. The main differences between FAT and the two coherent schemes are given as follows. First, compared with analog MIMO, FAT allows CSI-free transmission. Moreover, FAT reduces transmission latency significantly by using linear analog modulation while quantization is needed in digital MIMO, which will extend the total transmission period.

**Performance evaluation**: The latency and learning performances of FAT are evaluated in Fig. 4 (c) and Fig. 4 (d), respectively. While FAT is free of channel-training, benchmark schemes incur training overhead that can be quantified by the fraction of a frame allocated for the purpose i.e., the ratio $P/(P+D)$ with $P$ and $D$ denoting the pilot duration and payload data duration, respectively. Given the classification-error rate of $10^{-3}$, the curves of overhead versus Doppler shift for the FAT and two mentioned benchmarking schemes are displayed in Fig. 4 (c). One can observe that the overhead grows monotonically with the Doppler shift as the channel fading becomes faster. For high-mobility with Doppler approaching $10^{-2}$, the overhead can be more than 12% and 6% for digital and analog coherent MIMO, respectively. Furthermore, given the same performance, digital coherent MIMO (with QPSK modulation and 8-bit quantization) requires 4 times more frames for transmitting the training dataset than the two analog schemes. In addition, classification error rates of different schemes are compared in Fig. 4 (d) by varying Doppler shift. It is observed that in the range of moderate to large Doppler shift (i.e., larger than $6\times10^{-3}$), the proposed FAT outperforms the benchmarking schemes. The above observations suggest that the GAE-based analog transmission can support fast data acquisition in edge learning with a guaranteed performance.

### c) Research Opportunities

Learning-driven signal encoding is an important research direction in the field of edge learning. Some research opportunities are described as follows.

- **Gradient-data encoding**. For the trainings of AI models at the edge, especially in the setting of *federated learning*, the transmission of stochastic gradients from edge devices to the edge server lies at the center of the whole edge learning process. However, the computed gradients may have a high dimensionality, which is extremely communication inefficient. Fortunately, it is found in the literature that, by exploiting the inherent sparsity structure, a gradient for updating an AI model can be truncated appropriately without significantly degrading the training performance [9]. This inspires the design of gradient compression techniques to reduce communication overhead and latency.

- **Motion-data encoding**. A motion can be represented by a sequence of subspaces, which is translated into a trajectory on a Grassmann manifold. How to encode a motion dataset for both efficient communication and machine learning is an interesting topic for edge learning. For example, relevant designs can be built on the GAE method.

- **Channel-aware feature-extraction**. Traditionally, to cope with hostile wireless fading channels, various signal processing techniques such as MIMO beamforming and adaptive power control are developed. The channel-aware signal processing can be also jointly designed with the feature extraction in edge learning systems. Particularly, recent research in [10] has shown the inherent analogy between the feature extraction process for classification and the non-coherent communication. This suggests the possibility to exploit the channel characteristics for efficient feature-extracting, giving rise to a new research area of channel-aware feature-extraction.

## V. Edge Learning Deployment

The success of edge learning depends on its practical deployment, which will be facilitated by recent advancements in a few key supporting techniques. Specifically, the thriving AI chips and software platforms lay the physical foundations for edge learning. Meanwhile, the recent maturity of MEC, supported by the upcoming 5G networks, provides a practical and scalable network architecture for implementing edge learning.

### a) AI Chips

Training an AI model requires computation- and data-intensive processing. Unfortunately, CPUs that have dominated computing in the last few decades fall short in these aspects, mainly for two reasons. First, the Moors's Law that governs the advancement of CPUs appears to be unsustainable, as transistor densification is soon reaching its physical limits. Second, the CPU architecture is not designed for number crunching. In particular, a small number of cores in a typical CPU cannot support the aggressive parallel computing, while placing cores and memory in separate regimes results in significant latency in data fetching. The limitations of CPUs have recently driven the semiconductor industry to design chips customized for AI. There exist diversified architectures for AI chips, many of which share two common features that are crucial for number crunching in machine learning. First, an AI chip usually comprises many mini-cores for enabling parallel computing. The number ranges from dozens for device-grade chips (e.g., 20 for Huawei Kirin 970) to thousands for server-grade chips (e.g., 5760 for Nvidia Tesla V100) [11]. Second, in an AI chip, memory is distributed and placed right next to mini-cores so as to accelerate data fetching. The rapid advancements in AI chips will provide powerful brains for fast edge learning.

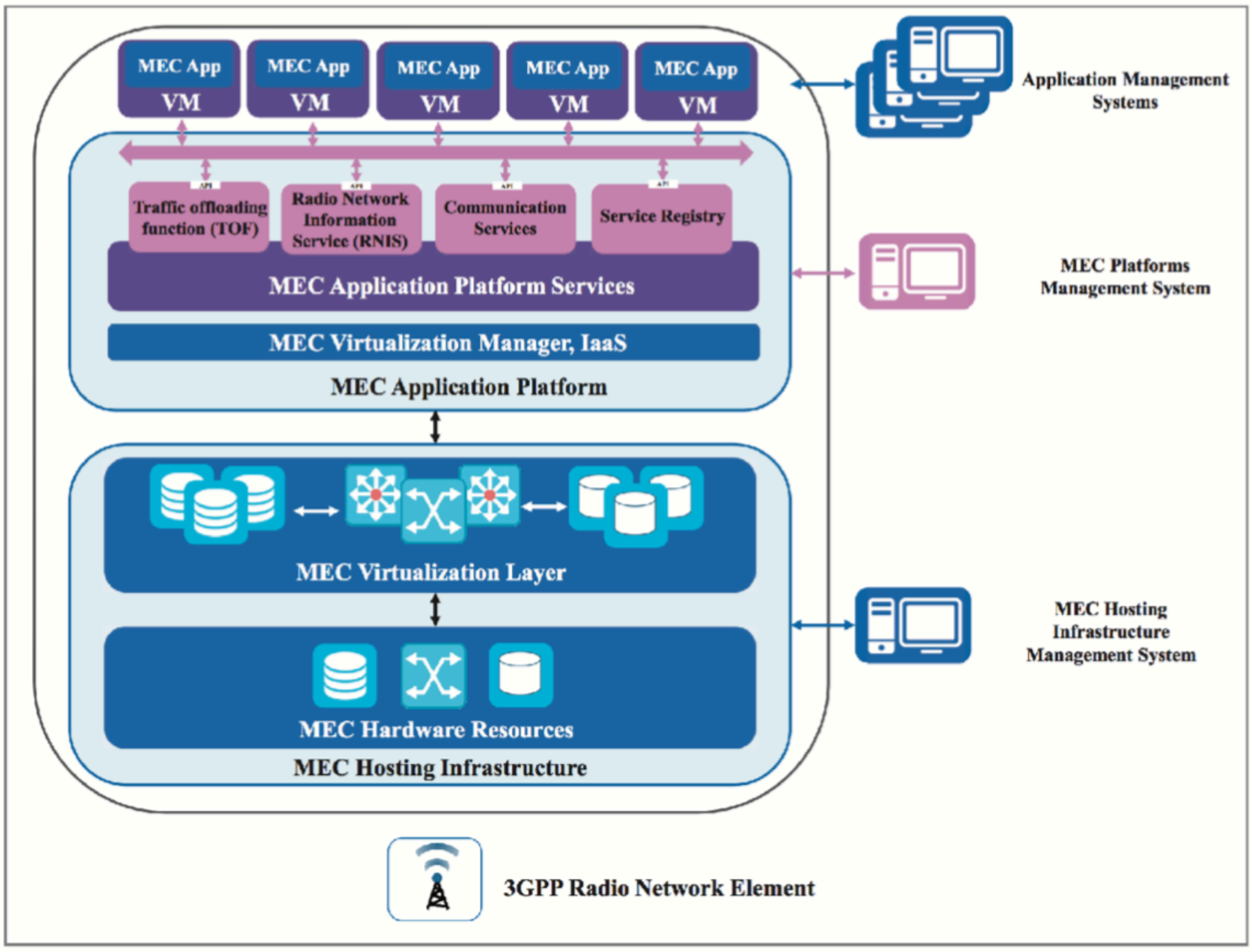

Fig. 5. 3GPP network virtualization architecture supporting implementation of edge computing and learning.

### b) AI Software Platform

Given the shared vision of realizing an intelligent edge, leading Internet companies are developing software platforms to provide AI and cloud-computing services to edge devices. They include AWS Greengrass by Amazon, Azure IoT Edge by Microsoft, and Cloud IoT Edge by Google. These platforms currently rely on powerful data centers. In the near future, as AI-enabled mission-critical applications become more common, these platforms will be deployed at the edge to implement edge learning. Most recently, two companies, Marvell and Pixeom, have demonstrated the deployment of Google TensorFlow micro-services at the edge to enable a number of applications, including objective detection, facial recognition, text reading, and intelligent notifications. To unleash the full potential of edge learning, we expect to see close cooperation between Internet companies and telecom operators to develop a highly efficient air interface for edge learning.

### c) MEC and 5G Network Architecture

First, the network virtualization architecture, standardized by 3GPP for 5G, as shown in Fig. 5, provides a platform for implementing edge computing and learning. The virtualization layer in the architecture aggregates all geographically distributed computation resources and presents them as a single cloud for use by applications in the upper layer. Different applications share the aggregated computation resources via virtual machines (VMs). Second, network function virtualization specified in the 5G standard enables the telecommunication operators to implement network functions as software components, achieving flexibility and scalability. Among others, the network functions support control and massage passing to facilitate selecting user-plane functions, traffic routing, computation resource allocation, and supporting mobility [12]. Last, VMs provide an effective mechanism for multiple edge-learning applications hosted at different serves to share the functions and resources of the same physical machines (e.g., operating systems, CPUs, memory, and storage).

## VI. Concluding Remarks

Edge learning, sitting at the intersection of wireless communication and machine leaning, enables promising AI-powered applications, and brings new research opportunities. The main aim of this article is to introduce a set of new design principles to the wireless communication community for the upcoming era of edge intelligence. The introduced learning-driven communication techniques can break the communication latency bottleneck and lead to fast edge learning, which are illustrated in three key topics: computation-oriented multiple access for ultra-fast data aggregation, importance-aware resource allocation for agile intelligence acquisition, and learning-driven signal encoding for high-speed data-feature transmission.

Besides the three presented research directions, there are many other research opportunities which deserve further exploration. Some of them are described as follows.

- **Is noise foe or friend?** In conventional wireless communication, noise is considered as the key obstacle to reliable communication. Thus the main focus has been on noise-mitigation techniques such as channel coding, diversity combining, and adaptive modulation. On the other hand, in machine learning, noise is not always harmful and can even be exploited for learning performance enhancement. For example, recent research shows that injecting noise perturbation into the model gradient during training can help loss-function optimization by preventing the learnt model being trapped at the poor local optimums and saddle points [13]. As another example, perturbing the training examples by a certain level of noise can be beneficial as it prevents the learnt model from overfitting to the training set, and thus endow on the model better generalization capability [14]. This motivates rethinking of the role of channel noise in edge learning. Apparently, the overuse of the conventional anti-noise techniques may lead to inefficient utilization of radio resources and even suboptimal learning performance. Therefore, how to regulate the channel noise to be at a beneficial level is an interesting topic in the area of learning-driven communication.
- **Mobility management in edge learning**. In edge learning, the connection between the participating devices and the edge server is transient and intermittent due to the mobility of device owners. This poses great challenges for realizing low-latency learning. Specifically, edge learning is typically implemented under the heterogeneous network architecture comprising macro and small-cell base stations and WiFi access points. Thus, users' movement will incur frequent handovers among the small-coverage edge servers, which is highly inefficient as excessive signalling overhead will arise from the adaptation to the diverse system configurations and user-server association policies. Moreover, the accompanying learning task migration will significantly slow down model training. As a result, intelligent mobility management is imperative for practical implementation of edge learning. The key challenge lies in the joint consideration of both the link reliability and task migration cost in the handover decision making.
- **Collaboration between cloud and edge learning**. Cloud learning and edge learning can complement each other with their own strengths. The federation between them allows the training of more comprehensive AI models that consist of different levels of intelligence. For example, in the industrial control application, an edge server can be responsible to the training of low-level intelligence such as anomaly detector, for tactile response to the environment dynamics. On the other hand, a cloud server can concentrate on crystallizing the higher-level intelligence, such as the regulating physical rules behind the observations, for a better prediction of the ambient environment. More importantly, the collaboration between the cloud and edge learning can lead to mutual performance enhancement. Particularly, the performance of the low-level edge AI can be fed back to the cloud as a learning input for continuously refining the high-level cloud AI. In return, the more accurate cloud AI can better guide the model training at the edge. Nevertheless, how to develop an efficient cooperation framework with minimum information exchange between the edge server and cloud server is the core challenge to be addressed.